\documentclass[twocolumn]{jpsj2} %% two-column layout
\title{Field-Induced Quasiparticle Excitation in Ca(Al$_{0.5}$Si$_{0.5}$)$_2$: Evidence for Unconventional Superconductivity}
\author{
Sogo {\sc Kuroiwa}$^{1}$, Hiroyuki {\sc Takagiwa}$^{1}$, Maki {\sc Yamazawa}$^{1}$, Jun {\sc Akimitsu}$^{1}$,\\
Kazuki {\sc Ohishi}$^{2}$, Akihiro {\sc Koda}$^{2}$, Wataru {\sc Higemoto}$^{2}$ and Ryosuke {\sc Kadono}$^{2}$\footnote{Also at School of Mathematical and Physical Science,
The Graduate University for Advanced Studies}}
\inst{$^{1}$Department of Physics, Aoyama-Gakuin University,
Fuchinobe, 5-10-1, Sagamihara, Kanagawa, 229-8558 \\
$^{2}$Institute of Materials Structure Science, High Energy Accelerator Research Organization (KEK), Tsukuba, Ibaraki 305-0801}

\recdate{February 10, 2004}
\abst{
The temperature ($T$) and magnetic field ($H$) dependence of the magnetic penetration depth, $\lambda(T,H)$, in Ca(Al$_{0.5}$Si$_{0.5}$)$_2$ exhibits significant deviation from that expected for conventional BCS superconductors. In particular, it is inferred from a field dependence of $\lambda(H)$ ($\propto H$) at 2.0 K that the quasiparticle excitation is strongly enhanced by the Doppler shift. This suggests that the superconducting order parameter in Ca(Al$_{0.5}$Si$_{0.5}$)$_2$ is characterized by a small energy scale $\Delta_S/k_B\le 2$ K originating either from anisotropy or multi-gap structure.
}

\kword{Ca(Al$_{0.5}$Si$_{0.5}$)$_2$, superconductivity, penetration depth, flux line lattice, $\mu$SR,\\}

\begin{document}
\sloppy
\maketitle
\par
Since the discovery of binary intermetallic compound superconductor MgB$_2$ ($T_{\rm c}$=39 K),\cite{mgb2} the origin of pair correlation leading to such high critical temperature has been drawing much interest.  Up to now, a majority of experiments as well as theories suggests that the pair correlation is mediated by a strong electron-phonon interaction, where the relatively high $T_{\rm c}$ is explained by the light mass of the two-dimensional honeycomb layer formed by B atoms.  The two-dimensional feature is also preferable for the strong electron-phonon interaction, and thereby the AlB$_2$-type crystal structure
common to MgB$_2$ is currently attracting attention as a possible basis for
developing new superconductors.
\par
In binary silicides, ThSi$_2$,\cite{thsi2} USi$_2$,\cite{usi2} and several rare-earth metal disilicides have the AlB$_2$-type structure. $\beta$-ThSi$_2$ is known to be a superconductor with a critical temperature $T_{\rm c}$=2.41 K.\cite{bthsi2} Above 16 GPa, CaSi$_2$ takes a AlB$_2$-like structure, and it becomes superconducting with $T_{\rm c}$=14 K.\cite{casi2} Recently, a new ternary silicide, Sr(Ga,Si)$_2$, which also has the AlB$_2$-type structure, was reported to be a superconductor with $T_{\rm c}$=3.4 K,\cite{srgasi} stimulating active investigation of analogous compounds. In this class of materials, ternary silicide Ca(Al$_{0.5}$Si$_{0.5}$)$_2$ has the highest critical temperature $T_{\rm c}$=7.7 K.\cite{caalsi} It is reported on this compound that the behavior of electron-heat capacity deviates from that of the BCS-type, and that the effect of hydrostatic pressure on $T_{\rm c}$ is positive.\cite{heat}  Unfortunately, despite various experiments so far, there is very little known on the structure of superconducting order parameter in Ca(Al$_{0.5}$Si$_{0.5}$)$_2$.
\par
In this paper, we report the quasiparticle excitation in the flux line lattice (FLL) state of Ca(Al$_{0.5}$Si$_{0.5}$)$_2$ studied by muon spin rotation ($\mu$SR). The magnetic penetration depth $\lambda$, which reflects the quasiparticle excitation, can be determined by $\mu$SR due to the inhomogeneity of the magnetic field distribution in the FLL state. We show the temperature ($T$) and magnetic field ($H$) dependence of $\lambda$ in polycrystalline sample of Ca(Al$_{0.5}$Si$_{0.5}$)$_2$.  Our result indicates that $\lambda$ is proportional to $T^3$ at low temperature, and that it exhibits a steep increase with almost linear dependence on the applied magnetic field. These findings strongly suggest that Ca(Al$_{0.5}$Si$_{0.5}$)$_2$ has an anisotropic order parameter or a multi-gap structure with a band having a small gap energy, which is qualitatively similar to the case in MgB$_2$.
\par
The samples were prepared by arc melting method with a stoichiometric mixture of Ca (99.9$\%$), Al (99.99$\%$), and Si (99.999$\%$)  (in 1:1:1 composition) under an Ar atmosphere. The structure of polycrystalline sample was examined using a powder x-ray diffraction technique.  Diffraction signal from Ca(Al$_{0.5}$Si$_{0.5}$)$_2$ phase was observed as a main contribution besides those from a small amount of impurity phases. Magnetic susceptibility and electrical resistivity measurements were performed with the SQUID magnetometer (MPMSR2) and the PPMS system (Quantum Design Co. Ltd.).  The residual resistivity ratio was estimated to be 2.85.

\begin{figure}[b]
\begin{center} \includegraphics[width=0.8\linewidth]{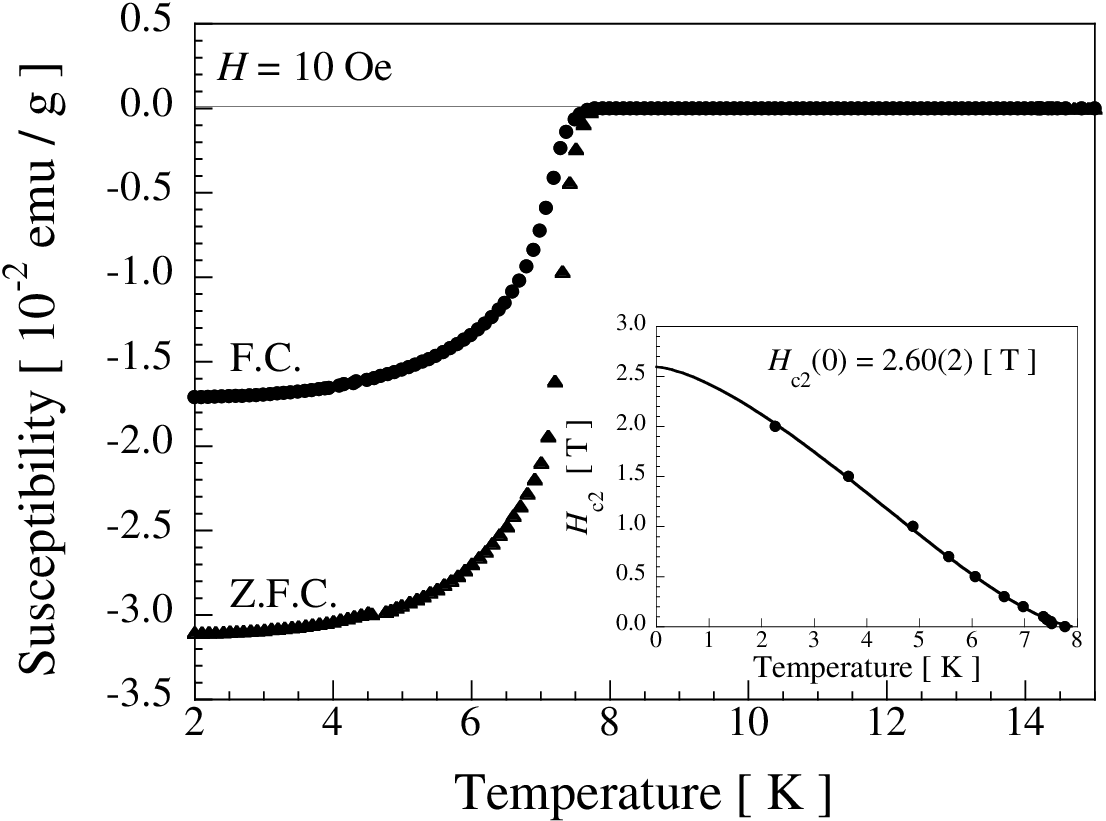}
\caption{Temperature dependence of magnetic susceptibility at 10 Oe. The inset shows the upper critical field versus temperature.} \label{fig1}
\end{center}
\end{figure}

Figure \ref{fig1} shows the temperature dependence of magnetic susceptibility at 10 Oe. The solid triangles and circles show data obtained in zero-field cooling (ZFC) and in field cooling (FC), respectively. In FC, a Meissner effect can be seen below 7.7 K. The inset of Figure \ref{fig1} shows the temperature dependence of the upper critical field ($H_{\rm c2}$) which was estimated from the temperature dependence of electrical resistivity at respective magnetic fields. The solid line is the result of fitting analysis by the following equation from a local-paring theory,
\begin{equation}
H_{\rm c2}(T)=H_{\rm c2}(0)(1-\tau^{3/2})^{3/2},\label{hc2}
\end{equation}
where $\tau=T/T_{\rm c}$, which yields $H_{\rm c2}(0)=2.60(2)$ T.
\par
The $\mu$SR experiment was performed on the M15 beam line at the Tri-University Meson Facility (TRIUMF, Canada) which provides a muon beam with a momentum of 29MeV$/$c. The specimen having a dimension of about 7$\times$7 mm$^2$ was mounted on sample holder and placed in a cryostat. The sample was field-cooled at every magnetic field points to minimize the effect of flux pinning. The temperature and magnetic field dependence of transverse field (TF-) $\mu$SR spectrum was obtained at $H$ = 0.05 T and $T$ = 2.0 K (=0.26$T_{\rm c}$), respectively.  Since the muon stops randomly on the length scale of the FLL, the muon spin precession signal $\hat{P}(t)$ provides a random sampling of the internal field distribution $B(\hat{r})$
\begin{equation}
\hat{P}(t){\equiv}P_x(t)+iP_y(t)=\int_{-\infty}^{\infty}n(B)\exp(i\gamma_\mu Bt+\phi)dB,
\end{equation}
\begin{equation}
n(B)=\langle\delta(B(\hat{r})-B)\rangle_r,
\end{equation}
where $\gamma_\mu$ is the muon gyromagnetic ratio (= 2$\pi$~135.53 MHz$/$T), $\phi$ is the initial phase of rotation, and $n(B)$ is the spectral density for the muon spin precession signal determined by the local magnetic field distribution. These equations indicate that the real amplitude of the Fourier transformed muon spin precession signal corresponds to the local field distribution $n(B)$. The London penetration depth in the FLL state is related to the second moment of the field distribution $\langle({\Delta}B)^2\rangle$, which is reflected in the $\mu$SR line shape.\cite{fll} For polycrystalline samples, the Gaussian distribution of local field is a good approximation,
\begin{eqnarray}
\hat{P}(t)&\simeq&\exp(-{\sigma}^2t^2/2)\exp(i\gamma_\mu Ht),\\
\sigma&=&\gamma_\mu\sqrt{\langle(\Delta B)^2\rangle}.
\end{eqnarray}
For the ideal triangular FLL with isotropic effective carrier mass $m^{\ast}$, $\lambda$ is given by the following relation,\cite{fll,penet2,penet3}
%\begin{equation}
%\sigma[\mu{\rm s^{-1}}]=4.83\times 10^4(1-h){\lambda}^{-2}[{\rm nm}],
%\end{equation}
%where $h=H/H_{\rm {c2}}(T)$, $\sigma$ is the muon spin relaxation rate.
%While the above form is useful for $h<0.25$ and $h>0.7$, we have a better
%approximation for arbitrary field,
\begin{equation}
\sigma[\mu{\rm s^{-1}}]=4.83\times 10^4(1-h)[1+3.9(1-h)^2]^{1/2}\lambda^{-2}[{\rm nm}].\label{sgmh}
\end{equation}
where $h=H/H_{\rm {c2}}(T)$, and $\lambda$ is given by the following relation of superconductive carrier density $n_s(T,H)$
\begin{equation}
\lambda^2=\frac{m^{\ast}c^2}{4{\pi}n_s(T,H)e^2}\label{lmdsq}
\end{equation}
It should be noted that the reduction of $\sigma$ with increasing field described by Eq.~(\ref{sgmh}) is due to the stronger overlap of $B(\hat{r})$ around vortices at higher fields while $\lambda$ is a constant.   The reduction of $n_s(T,H)$ due to the quasiparticle excitation leads to the enhancement of $\lambda$, which causes further reduction of $\sigma$ from that expected from Eq.~(\ref{sgmh}).

\begin{figure}[t]
\begin{center} \includegraphics[width=0.7\linewidth]{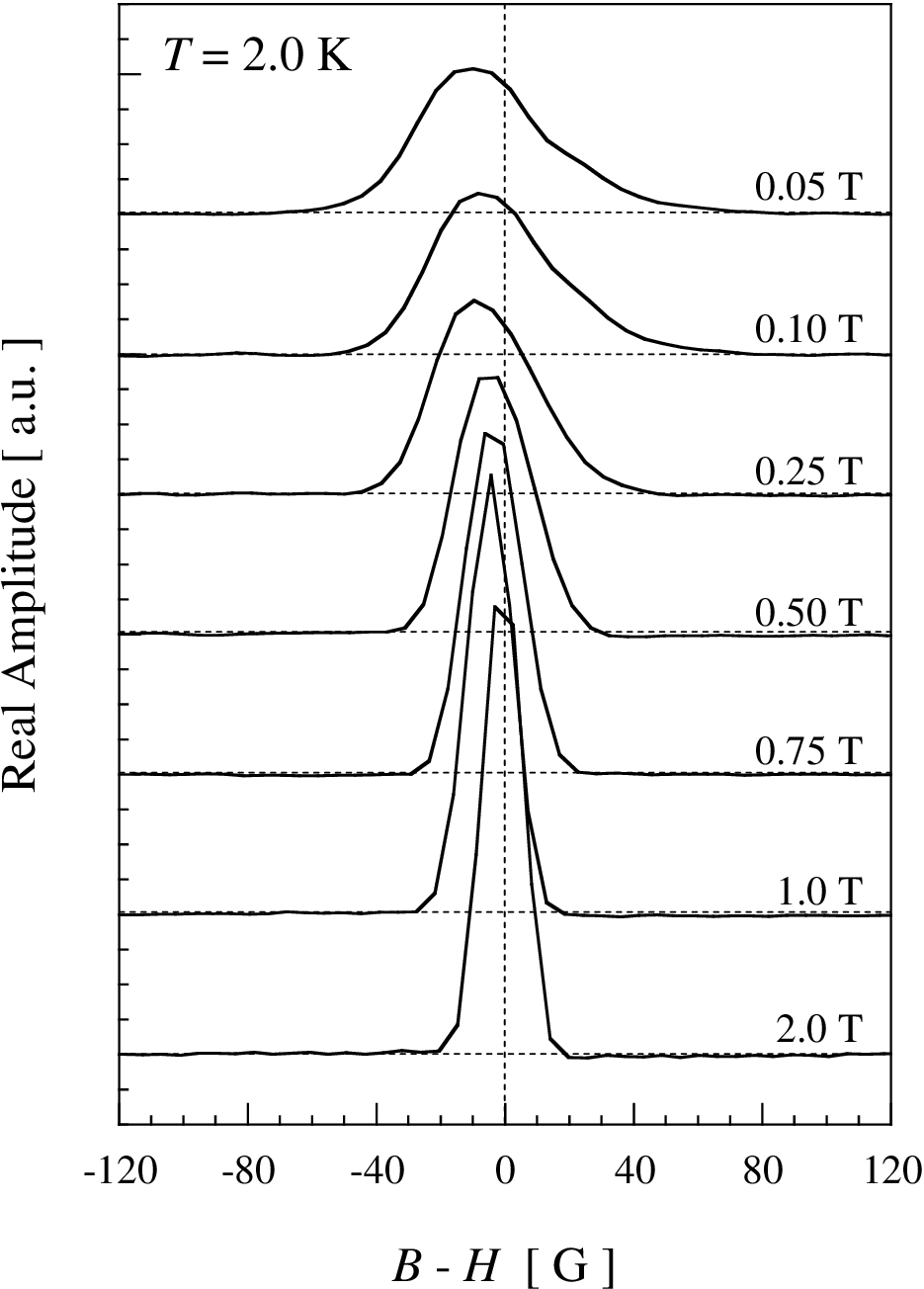}
\caption{Fast Fourier transform of $\mu$SR spectra in Ca(Al$_{0.5}$Si$_{0.5}$)$_2$ at 2.0 K under several applied magnetic fields.} \label{fig2}
\end{center}
\end{figure}

\par
Figure \ref{fig2} shows the fast Fourier transforms (FFT) spectrum of muon spin precession signal in Ca(Al$_{0.5}$Si$_{0.5}$)$_2$ for several magnetic fields at 2.0 K, where the real amplitude of FFT corresponds to the internal field distribution $n(B)$ in the FLL state convoluted with additional damping due to small random field from nuclear moments. As most explicit in FFT spectrum at 0.05 T, the line shape is characterized by the shift of peak to a lower field and also associated broadening of linewidth due to the formation of FLL.
\par
Considering possible contribution from muons stopping in the normal part of the specimen, we adopted two components with the following empirical form in analyzing the $\mu$SR time spectra,
\begin{equation}
A\hat{P}(t)=\sum_{i=1}^2A_i\exp\left(-\frac{\sigma_i^2t^2}{2}\right)\exp(\gamma_\mu H_i t+\phi_i)
\end{equation}
where the index $i$ refers to the components of superconducting ($i=1$) and normal ($i=2$) domains, $A$ is the total positron  decay asymmetry with $A_i$ being the partial asymmetry, $\sigma_i$ is the muon spin relaxation rate, $H_i$ is the central frequency, $\phi_i$ is the initial phase for respective components. The superconducting volume fraction ($=A_1/(A_1+A_2$) was obtained to be $\simeq0.95$.

\begin{figure}[t]
\begin{center} \includegraphics[width=0.9\linewidth]{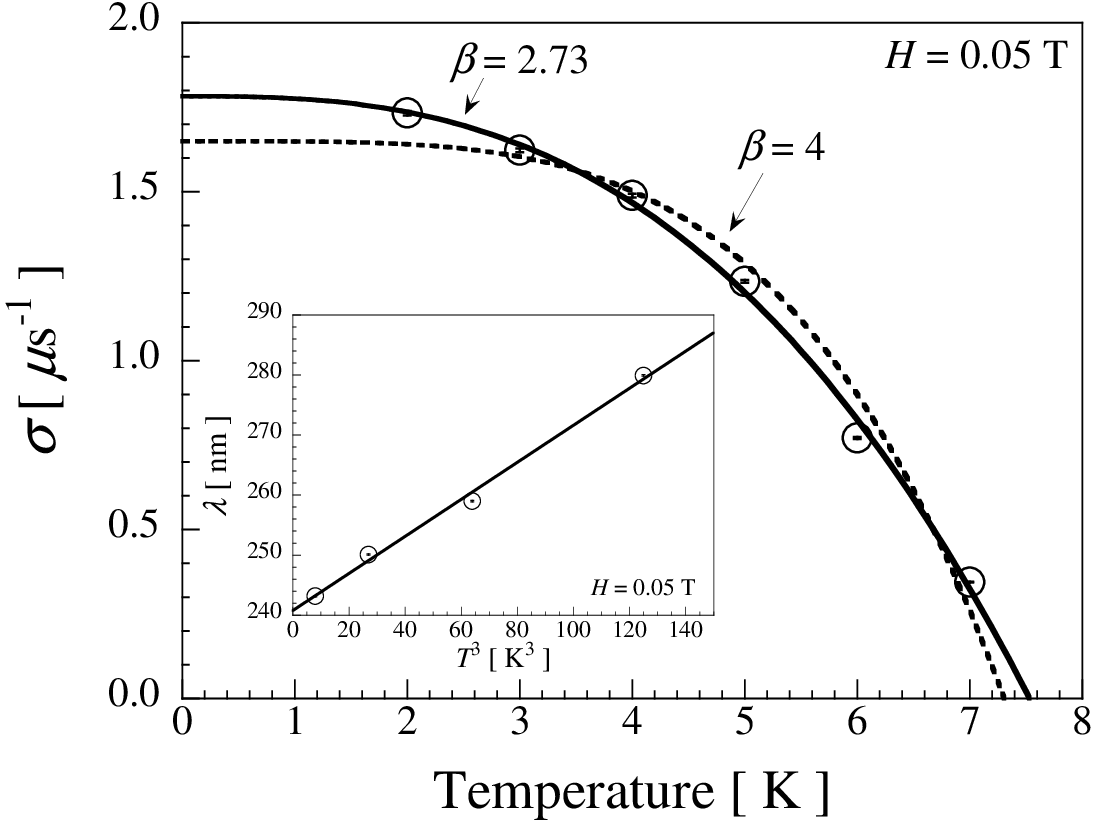}
\caption{Temperature dependence of the Gaussian relaxation rate $\sigma_1$ at 0.05 T.  The solid curve is a fitting result with $\sigma_1(T)=\sigma(0)[1-(T/T_{\rm c})^{\beta}]$, while the dashed curve corresponds to the case $\beta=4$. Inset: $\lambda$ plotted against $T^3$.} \label{fig3}
\end{center}
\end{figure}

\par
The temperature dependence of $\sigma_1\:(\propto n_s)$ is shown in Figure \ref{fig3}. In the region below $T_{\rm c}$ where the FLL is formed, $\sigma_1$ increases with decreasing temperature. According to the empirical two-fluid model (which is a good approximation of the BCS theory), the following relation is expected to hold,
\begin{equation}
\sigma_1(T)=\sigma_1(0)[1-\tau^4]\propto n_s(T)\propto \lambda^{-2}(T).
\end{equation}
Solid line is the result of fitting analysis by a similar formula with an arbitrary power,
\begin{equation}
\sigma_1(T)=\sigma_1(0)[1-\tau^{\beta}],
\end{equation}
whch yields $\beta = 2.73$ and $T_{\rm c}= 7.53(1)$ K when both $\beta$ and $T_{\rm c}$ are assumed to be free parameters. The result also means that the deviation of $\lambda$,
\begin{equation}
\delta\lambda=\lambda(T)-\lambda(0),
\end{equation}
exhibits a tendency predicted for the case of line nodes ($d$-wave pairing) with some disorder (i.e., dirty limit, where $\delta\lambda\propto T^2$).\cite{delta} Since the gap energy $\Delta(T)\simeq\Delta_0(1-\tau^4)$ is predicted to be least dependent on temperature for $\tau\le 0.4$ in the BCS model, the observed deviation of $\beta$ from 4 suggests the presence of excess quasiparticle excitation.  This leads to the possibility that the order parameter has an anisotropic strutcure, where the excess quasiparticles are induced at the vicinity of nodes ($|\Delta_{\hat{k}}|/ k_B\le 2$ K). Another possibility may be that, assuming a multi-gap structure with $\Delta_S$ being one of the smallest gap energy, the excess quasiparticles are due to the thermal activation over $\Delta_S$  ($\Delta_S/ k_B\le 2$ K) ; unfortunately, the absence of data below 2.0 K did not allow us to perform the reliable fitting analysis using a two-gap model\cite{Bouquet:01}.
\par
\begin{figure}[t]
\begin{center} \includegraphics[width=0.9\linewidth]{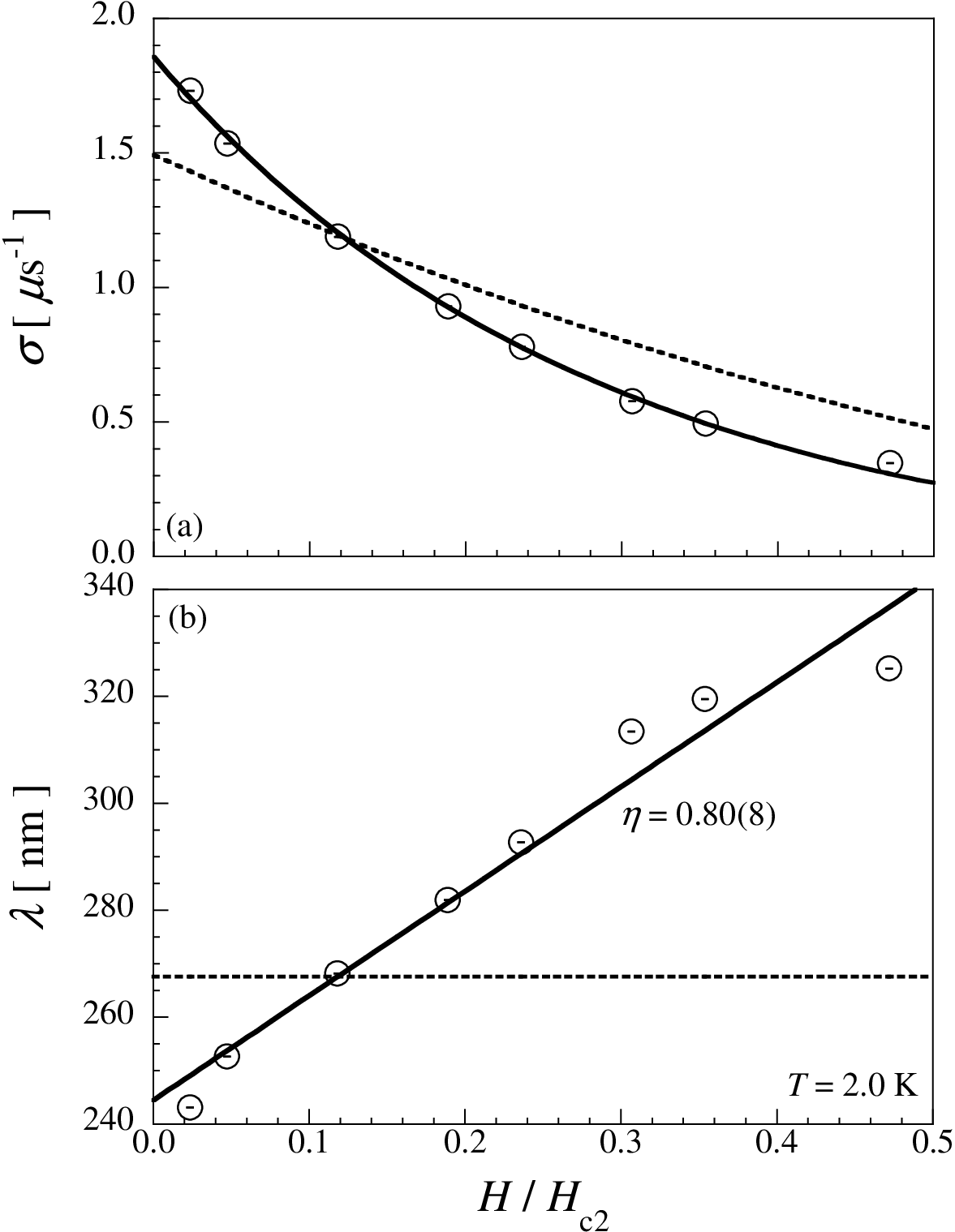}
\caption{Magnetic field dependence of the Gaussian relaxation rate $\sigma_1$ (a), and penetration depth $\lambda$ (b) at 2.0 K. Solid curves are fitting result with $\lambda(h)=\lambda(0)[1+{\eta}h]$, and the dashed curves correspond to $\eta=0$.} \label{fig4}
\end{center}
\end{figure}

The presence of an energy scale smaller than that for the single BCS-gap energy $\Delta_0$ is further supported by the magnetic field dependence of $\lambda$.  As shown in Fig.~\ref{fig4}a, $\sigma(h)$ decreases with increasing field much steeply than expected by Eq.~(\ref{sgmh}). Accordingly, $\lambda$ exhibits a strong field dependence, where $\lambda$ increases almost linearly with $h$ (see Fig.~\ref{fig4}b). This is similar to the case of NbSe$_2$\cite{nbse2} and YNi$_2$B$_2$C\cite{ynbc} which exhibit characters specific to anisotropic order parameters in spite of the suggested $s$-wave symmetry, or to that of MgB$_2$\cite{Ohishi:03} having a two-gap structure. Solid curves in Fig.~\ref{fig4} is the result of fitting by the following linear relation,
\begin{equation}
\lambda(h)=\lambda(0)[1+{\eta}h],\label{lmdeh}
\end{equation}
where $\eta$ is a dimension-less parameter which represents the strength of pair breaking effect.  Fitting yields $\eta$ = 0.80(8) (with $H_{\rm c2}=2.12(2)$ T determined by Eq.~(\ref{hc2})) which is comparable with that in NbSe$_2$, YNi$_2$B$_2$C and MgB$_2$ ( i.e., $\eta$ = 1.61,
0.97, and 1.27, respectively).

\par
In the FLL state, the quasiparticle energy spectrum around the vortex cores
is shifted by an amount
$\varepsilon=\hat{p}\cdot\hat{v}_s$ due to the semiclassical Doppler shift,
where $\hat{p}$ is the quasiparticle momentum ($\simeq m^*\hat{v}_F$), and $\hat{v}_s$ is the velocity of supercurrent.  Since the density of state (DOS), $N(E)$, for anisotropic order parameters takes a non-zero value for an energy $E>0$ , quasiparticles can be excited by the Doppler shift outside of the vortex cores with a population proportional to $N(\varepsilon)$, leading to the enhancement of $\lambda$.\cite{Volovik}  The magnitude of $\eta$ represents the degree for the increase of DOS for quasiparticles, which must be roughly proportional to the phase volume of the Fermi surface where the Doppler shift exceeds the gap energy ($\varepsilon>|\Delta_{\hat{k}}|$).  It also follows that the effect depends on the direction of $\hat{v}_s$ (and hence that of the external field ${\bf H}$ relative to the order parameter) in a single crystalline specimen.  According to Volovik, the quasiparticle density of state for anisotropic order parameter is
\begin{equation}
N_{\rm deloc}(0)\simeq N_{\rm F} K \xi_{\rm GL}^2\sqrt{h}
\equiv N_{\rm F}g(h),\label{gh}
\end{equation}
\begin{equation}
K\propto\int_{|\Delta_{\hat{k}}|<\varepsilon}|\Delta_{\hat{k}}|d{\bf k},
\end{equation}
where $N_{\rm F}$ is the DOS for the normal state
and $K$ is a constant on the order of unity\cite{Volovik}.
It is important to note that $K$ is proportional to the phase volume
of the low excitation energy in $\Delta_{\hat{k}}$, thereby carrying information
on the degree of anisotropy for $\Delta_{\hat{k}}$; the factor $h^{1/2}$
comes from the inter-vortex distance ($\propto h^{-1/2}$) multiplied by
the number of vortices ($\propto h$).
The superfluid density at a given field is then
\begin{equation}
n_{\rm s}(h)\simeq n_{\rm s}(0)[1-g(h)],
\end{equation}
which is directly reflected in the magnetic penetration depth through Eq.~(\ref{lmdsq}).
Therefore, as a mean approximation, we have
\begin{equation}
\lambda(h)=\frac{\lambda(0)}{\sqrt{1-g(h)}}
\sim\lambda(0)[1+cK\xi_{\rm GL}^2h],
\label{lmdh}
\end{equation}
where $c\simeq1.5$ for $0<h<0.5$.\cite{Kadono:04}
Thus, the comparison between Eqs.~(\ref{lmdeh}) and (\ref{lmdh}) yields
\begin{equation}
\eta\simeq cK\xi_{\rm GL}^2,
\end{equation}
indicating that the slope $\eta$ reflects the phase volume of
the Fermi surface where $|\Delta_{\hat{k}}|<\varepsilon$.
The situation in the case of multi-gap state is similar when $\Delta_S\le k_BT$.  For example in the two-gap model\cite{Bouquet:01}, using the relative weight, $1-x$, of the quasiparticle DOS for the smaller gap $\Delta_S$ ,  we have
\begin{equation}
K\propto 1-x.
\end{equation}

Since the Doppler shift is far smaller than the gap energy in usual situation for the isotropic gap, no such enhancement is expected for the conventional $s$-wave pairing (i.e., $\eta\ll$1 for $|\Delta_0|,\;|\Delta_S|\gg k_BT$). For example in Y(Ni$_{0.8}$Pt$_{0.2})_2$B$_2$C, which behaves as a conventional BCS superconductor with an isotropic gap, it is reported that $\eta\simeq0$.\cite{ynpbc} This is in good contrast with the case of YNi$_2$B$_2$C in which an anisotropic gap (two-gap state with point nodes) is strongly suggested by other experiments.\cite{ynbc2,Izawa}. On the other hand, a stronger magnetic field dependence is predicted for the case of $d$-wave pairing due to the relatively large value of $K$. Typical examples of the $d$-wave pairing are found in high-$T_{\rm c}$ cuprates, in which $\eta$ is reported to be 5-6.6 for YBCO.\cite{ybco3}  The comparison of those earlier results with our result suggests that Ca(Al$_{0.5}$Si$_{0.5}$)$_2$ has an anisotropic order parameter comparable to that of YNi$_2$B$_2$C or a multi-gap order parameter in which one of those has a small gap energy ($|\Delta_S|/k_B\le 2$ K) as found in MgB$_2$.
\par
Unfortunately, our results were obtained using polycrystalline samples, which makes it difficult to deduce $\lambda$ by analyzing data using the well-defined microscopic model. Early experiments on high quality single crystal sample of Ca(Al$_{0.5}$Si$_{0.5}$)$_2$ have reported that the upper critical field has an anomalous angular dependence which deviates from the Ginsburg-Landau anisotropic mass model.\cite{upper}  Moreover, very recent report suggests that the crystal structure has clear fivefold and sixfold superlattice.  Therefore, we are preparing high quality single crystals of Ca(Al$_{0.5}$Si$_{0.5}$)$_2$ for further $\mu$SR study of this compound in more detail.

In conclusion, We have performed $\mu$SR experiments on polycrystalline
Ca(Al$_{0.5}$Si$_{0.5}$)$_2$ to elucidate the structure of
superconducting order parameter. We found that $n_s(t)$ exhibits a considerable
deviation from that expected for the BCS prediction, and that
$\lambda(h)$  exhibits significant increase with applied magnetic
field. These results strongly suggest the presence of unconventional
order parameter in Ca(Al$_{0.5}$Si$_{0.5}$)$_2$.

We thank all the TRIUMF $\mu$SR staff for their technical support. This
work was partially supported by a Grant-in-Aid for Science Research on
Priority Areas from the Ministry of Education, Culture, Sports, Science,
and Technology of Japan.

\end{document}